**A novel test for selection on *cis*-regulatory elements reveals positive and negative selection acting on mammalian transcriptional enhancers**

Justin D. Smith[1], Kimberly F. McManus[2], and Hunter B. Fraser[2]*

[1]Department of Genetics, and [2]Department of Biology, Stanford University, Stanford CA 94305.

*Correspondence: hbfraser@stanford.edu

**Abstract**

Measuring natural selection on genomic elements involved in the cis-regulation of gene expression—such as transcriptional enhancers and promoters—is critical for understanding the evolution of genomes, yet it remains a major challenge. Many studies have attempted to detect positive or negative selection in these noncoding elements by searching for those with the fastest or slowest rates of evolution, but this can be problematic. Here we introduce a new approach to this issue, and demonstrate its utility on three mammalian transcriptional enhancers. Using results from saturation mutagenesis studies of these enhancers, we classified all possible point mutations as up-regulating, down-regulating, or silent, and determined which of these mutations have occurred on each branch of a phylogeny. Applying a framework analogous to $K_a/K_s$ in protein-coding genes, we measured the strength of selection on up-regulating and down-regulating mutations, in specific branches as well as entire phylogenies. We discovered distinct modes of selection acting on different enhancers: while all three have experienced negative selection against down-regulating mutations, the selection pressures on up-regulating mutations vary. In one case we detected positive selection for up-regulation, while the other two had no detectable selection on up-regulating mutations. Our methodology is applicable to the growing number of saturation mutagenesis data sets, and provides a detailed picture of the mode and strength of natural selection acting on *cis*-regulatory elements.

**Introduction**

Noncoding regions comprise the vast majority of genomic regions under selective constraint in mammals (Lindblad-Toh et al. 2011), harbor most common genetic variants influencing human disease (Hindorff et al. 2009), and may be the source of most evolutionary adaptations (King and Wilson 1975; Jones et al. 2012; Fraser 2013). Yet our ability to measure natural selection in noncoding regions has lagged far behind our ability to do so in the small fraction of the genome that codes for protein (Zhen and Andolfatto 2012).

In order to infer natural selection, one must be able to reject a null model of neutral evolution (Kimura 1984). In protein-coding regions, a convenient proxy for neutral mutations is the synonymous mutations that do not alter the amino acid sequence. By comparing the rate of accumulating potentially functional nonsynonymous mutations (abbreviated as $K_a$) to the synonymous rate ($K_s$), selection can be inferred (Kimura 1977): a slower nonsynonymous rate ($K_a/K_s < 1$) reflects negative (purifying) selection against changing protein sequence; a faster rate ($K_a/K_s > 1$) reflects positive selection for changing protein sequence; and approximately equal rates ($K_a/K_s \approx 1$) means that selection cannot be inferred. Because synonymous and nonsynonymous sites are interdigitated within every protein-coding gene, their mutation rates should not differ greatly, facilitating a direct comparison of the two. This framework has proven useful for studying protein-coding regions, but $K_a/K_s$ cannot be calculated for noncoding regions.

Many studies of noncoding evolution have therefore taken alternative approaches, such as scanning multiple genomes for noncoding regions with unusually rapid evolutionary rates (Pollard et al. 2006; Prabhakar et al. 2006; Haygood et al. 2007; Kim and Pritchard 2007; Bird et al. 2007). For example, one of the first studies of rapid evolution in the human lineage discovered 49 "human accelerated regions" (HARs), which are enriched near genes involved in transcriptional regulation (Pollard et al. 2006). However this study illustrates a major caveat for such analyses: although accelerated divergence is typically attributed to positive selection, many HARs were subsequently shown to likely result from biased gene conversion (Dreszer et al. 2007; Galtier and Duret 2007), a process that can lead to rapid divergence in the complete absence of selection. This highlights the need for comparisons to neutrally evolving sites that are matched in mutation rate to each noncoding region of interest. Although many candidate sources of neutral sites have been proposed (e.g. nearby intronic sites, transposons, or synonymous sites), none of these is entirely neutral, and because they are all located outside of the noncoding elements being tested, regional differences in mutation rates or recombination rates are possible (Andolfatto 2008). As a result, there is still no consensus as to what constitutes a suitable reference for noncoding elements (Zhen and Andolfatto 2012).

Another issue with studies scanning genomes for regions of rapid divergence is that even when positive selection is acting on a cis-regulatory element (e.g. for higher transcriptional activity), there is likely to be negative selection simultaneously acting on the same element (e.g. purging mutations that disrupt the element's function) (Zhen and Andolfatto 2012). When negative selection dominates—as is likely to be required for almost any element to maintain its function—scanning the genome for rapid evolution will not detect the positive selection, because the rate of the entire element will not be faster than neutral. In other words, "averaging" across sites in a noncoding element that are subject to different selection pressures will reduce power to detect selection.

One potential solution is to separate sites within a cis-regulatory element into multiple classes, and compare evolutionary rates between classes. This has the potential to solve both of the issues discussed above: mutation rates should be similar if sites of different classes are interspersed, and power to detect selection will be maximized if sites within each class experience similar selection pressures. An example of this approach is a metric called $K_b/K_i$, which compares the evolutionary rate within known transcription factor binding sites (TFBS) ($K_b$) with that outside TFBS ($K_i$), with an excess of substitutions in TFBS ($K_b/K_i > 1$) suggesting positive selection (Hahn et al. 2004). Although this has advantages over measuring the overall evolutionary rate across an entire enhancer or promoter, it has several drawbacks. Perhaps the most important is that most naturally occurring genetic variants affecting transcription factor binding fall *outside* of any recognizable TFBS (Zheng et al. 2010; Kasowski et al. 2010). Not only are these effects not captured by $K_b$, they also can render $K_i$ an underestimate of the neutral evolutionary rate (when mutations outside of TFBS are under negative selection), leading to inflated $K_b/K_i$ and possibly false inference of positive selection. This same caveat applies to more recent studies of selection on TFBS as well (Arbiza et al 2013). Another method that takes into account the effects of substitutions on TFBS motif strengths, while not assuming that changes outside of TFBS are neutral, still does not account for the effects of functional substitutions outside of known TFBS (Moses 2009).

Our goal here was to design a robust framework for detecting natural selection in noncoding regions. To achieve this, we utilized data from saturation mutagenesis studies, which measure the effect of every possible single nucleotide variant (SNV) within specific cis-regulatory elements (Patwardhan et al. 2009; Melnikov et al. 2012; Patwardhan et al. 2012; Kwasnieski et al. 2012). In these studies, sequence constructs containing every SNV within a promoter or enhancer are produced, either by large-scale oligonucleotide synthesis (Patwardhan et al. 2009; Melnikov et al. 2012; Kwasnieski et al. 2012) or by traditional oligonucleotide synthesis with some degeneracy introducing random SNVs (Patwardhan et al. 2012). Their transcriptional outputs are then measured, most often using high-throughput RNA sequencing of short transcribed "barcodes" that uniquely identify each construct. This type of data allows the classification of all possible SNVs into one of three classes: up-regulating (i.e. increasing transcription over the reference enhancer sequence), down-regulating, or silent. We can then compare the rate of fixation of (for example) up-regulating SNVs with the silent ones that are likely under little or no selection. Because the three classes of sites are interdigitated with one another, differences in regional mutation rates are unlikely. Directly analogous to $K_a/K_s$, we call these metrics $K_u/K_n$ and $K_d/K_n$ for measuring selection on up-regulating and down-regulating substitutions, respectively. Perhaps the most important novel aspect of these metrics is that because the up- and down-regulating SNVs are assessed for selection separately, we can potentially detect positive selection on one class, even when negative selection on the other class would have masked any signal when considering the element as a whole.

As an initial proof-of-principle, we applied our approach to the three liver enhancers studied by Patwardhan et al. (2012). The three enhancers, named ALDOB, ECR11, and LTV1, are located within or proximal to the *ALDOB*, *DHRS9*, and *Zfp36* genes respectively (Gregori et al. 2002; Kim et al. 2011). *ALDOB* is a glycolytic enzyme, fructose-1,6-bisphosphate aldolase; *DHRS9* is a short chain alcohol dehydrogenase/reductase; and *Zfp36* is a zinc-finger RNA-binding protein that binds and degrades cytokine mRNAs. These three enhancers were dissected by measuring the transcriptional output *in vivo* of over 640,000 distinct mutant enhancers, differing on average from the wild-type by SNVs at 2.1%-3.1% of sites; each nucleotide within each enhancer was mutated, on average, in over 4,000 distinct constructs. This saturation mutagenesis allowed the robust empirical estimation of each SNV's effect on transcription, providing the input for our approach.

**Results**

Our methods for detecting natural selection on noncoding elements, $K_u/K_n$ and $K_d/K_n$, are outlined in Figure 1. Briefly, they compare the rate of up- or down-regulating substitution with the rate of silent substitution from the same enhancer element (see Materials and Methods). Values significantly greater than one indicate likely positive selection, whereas those less than one imply negative selection, directly analogous to the commonly used protein-coding metric $K_a/K_s$ (Kimura 1977). Because the selective regimes are inferred for up- and down-regulation separately, they may reflect entirely distinct modes of selection.

To avoid potential issues with distant comparisons (including less accurate ancestral reconstruction and epistatic interactions; see Discussion), we focused on species within the same phylogenetic order as the original mutagenized enhancer (rodents for LTV1, and primates for ALDOB and ECR11). We obtained orthologous sequences for the three enhancers, aligned them, and reconstructed ancestral sequences by maximum likelihood (see Materials and Methods). We then calculated $K_u/K_n$ and $K_d/K_n$ for each individual branch in the phylogeny, as well as for overall phylogenies.

For the LTV1 enhancer, we found an overall $K_u/K_n$ ratio of 1.45 (Fisher's Exact p = 1.5 x $10^{-5}$ for the null model of neutrality) and $K_d/K_n$ ratio of 0.51 (p = 1.2 x $10^{-9}$) among rodents (Figure 2). No individual branches had a significant $K_u/K_n$ ratio, but three individual branches (branches D, F, and M in Figure 2C) did have a significantly lower $K_d/K_n$ ratio than expected under neutrality. These results suggest that within rodents there was positive selection for up-regulating mutations in this enhancer, coupled with negative selection against down-regulating mutations.

For the ALDOB enhancer, we found an overall $K_u/K_n$ ratio of 1.14 (p = 0.56) and a $K_d/K_n$ ratio of 0.48 (p = 1.7 x $10^{-4}$) in primates (Figure 3). No individual branches had a significant $K_u/K_n$ ratio, while one branch (branch E in Figure 3C) had a significantly lower $K_d/K_n$ ratio than expected under neutrality. These results suggest that within primates there was negative selection against down-regulating mutations, but no detectable selection on up-regulating mutations in this enhancer.

For the ECR11 enhancer, we found an overall $K_u/K_n$ ratio of 0.95 (p = 0.90) and a $K_d/K_n$ ratio of 0.67 (p = 0.065) (Figure 4), with no individual branches reaching significance. Therefore we cannot reject the null hypothesis of neutral evolution for primates as a whole. However closer examination revealed that these results were primarily driven by just two branches (D and E, the lemur clade) which accounted for 65% of all down-regulating substitutions in primates, and were evolving with no detectable selection against down-regulating SNVs ($K_d/K_n$ values of 0.90 and 0.93; Figure 4c). Excluding the lemur branches resulted in an overall $K_d/K_n$ of 0.44, significantly lower than expected under neutrality (p = 0.018). In contrast, $K_u/K_n$ was similar within vs. outside the lemur clade (values of 0.998 and 0.91, respectively). This suggests that the ECR11 enhancer may have experienced selection against down-regulating SNVs specifically in simians (monkeys and apes), although the *ad hoc* nature of this analysis precludes a definitive conclusion.

To test the robustness of our results for all three enhancers, we repeated the overall-phylogeny $K_u/K_n$ and $K_d/K_n$ calculations for each enhancer with three variations. First, we excluded individual branches from the analysis (Supplemental Table 2), to test whether any single branches may have disproportionate effects. Only for ECR11 $K_d/K_n$ was there a mixture of nominally significant and non-significant results (primarily due to the lemur clade described above). Second, we tested the effect of excluding random subsets of sites within each enhancer from analysis, to determine whether a small number of outlier sites may be driving the results. Even with up to 75% of sites excluded from analysis, we do not see any bias towards increased or decreased $K_u/K_n$ or $K_d/K_n$, though power to detect selection does decrease (Supplemental Figure 1). Third, we varied the p-value cutoff at which we classified SNVs as silent or functional. At three different cutoffs, we observed very similar patterns of significance as in our original analysis (Supplemental Table 3). Together, these analyses suggest that our results are not driven by outlier branches, outlier sites, or the exact p-value cutoff used.

**Discussion**

We have developed a new framework for measuring the strength of natural selection acting on cis-regulatory elements. Taking advantage of massively parallel experimental measurement of the effects of every possible point mutation, our approach can reveal either positive or negative selection acting on either up- or down-regulating mutations. The three enhancers we used as a proof-of-principle have all experienced negative selection against down-regulating mutations in at least some lineages, and one (LTV1) has also been subject to positive selection for up-regulation. Although we do not know the reason for this positive selection, it may involve an advantage of lower levels of cytokines whose mRNAs are degraded by *Zfp36* (the target of LTV1) (Brahma et al 2012). It is interesting to note that in no case was the phylogeny-wide $K_u/K_n$ significantly less than one, while $K_d/K_n$ was significantly less than one for all three enhancers (in at least part of the phylogeny). This may reflect the tendency for over-expression to be less deleterious than under-expression (Sopko et al 2006). Re-examination of this pattern once data are available for more enhancers would be informative.

A key assumption of this approach is a minor role of epistasis, or context-dependence of SNVs. If epistasis between SNVs was widespread, we would have to measure the impact of each SNV in the precise genetic background in which it occurred, a considerably more challenging experiment than mutagenesis of a single enhancer. Epistasis can be quantified in these enhancers because the set of mutant enhancers tested for activity contained not only every possible SNV, but also over 99.999% of all possible *pairs* of variant sites, only ~0.1% of which exhibit pairwise epistasis in their effects on transcription (Patwardhan et al. 2012). Furthermore, even among the epistatic SNVs, only the minority that alter the SNVs' classification (as up-regulating, down-regulating, or silent)—as opposed to the magnitude of up- or down-regulation—would have any impact on our approach. The rarity of strong epistasis implies that most SNVs will likely have the same effect direction (e.g. up-regulating) whether occurring in the background of (for example) a human enhancer, or the same enhancer in an ancestral primate, and thus epistasis is unlikely to have a significant effect on the overall patterns of selection that we have inferred. However more experiments would have to be performed to establish this definitively.

Although our framework is based on the widely used $K_a/K_s$ approach for protein-coding sequences, it does not suffer from several important limitations of $K_a/K_s$. For example: 1) $K_a/K_s$ reflects an "average" selection pressure across sites; if both positive and negative selection are acting on the same protein, only the more dominant one (typically negative selection) will be apparent (while it is possible to estimate $K_a/K_s$ at single codons or to partition proteins into different selection classes, these approaches require very large numbers of aligned sequences and/or nontrivial assumptions about the distribution of $K_a/K_s$ values among classes (Yang and Bielawski 2000)). 2) Because we almost never know the functional impact of specific amino acid substitutions, a $K_a/K_s > 1$ is uninformative with respect to what trait natural selection is actually favoring (e.g. higher or lower activity of an enzyme). 3) Synonymous sites are not actually neutral; treating them as such, as is the common practice, inflates estimates of $K_a/K_s$ and can lead to spurious evidence of positive selection. This is not merely a theoretical concern, as synonymous sites have been found to be under negative selection in every species studied to date (Akashi 1994; Hirsh et al. 2005; Stoletzki and Eyre-Walker 2007; Plotkin and Kudla 2011; Lawrie et al 2013).

In contrast, our approach provides improvements in all three of these areas. Specifically: 1) Our metrics are able to detect both positive and negative selection acting simultaneously on different subsets of sites (as exemplified by LTV1), therefore avoiding much of the problem of "averaging" across sites suffered by $K_a/K_s$ applied to entire proteins. 2) Because we know the functional effects of each mutation on enhancer activity, we can infer not only the mode of selection (e.g. positive), but also what the selection is for (e.g. greater transcription). 3) Our metrics do not assume that a certain subset of sites (such as synonymous sites) are neutral, but instead rely on thousands of empirical measurements of the effect of every individual SNV.

It is also informative to compare our approach with previous studies that have scanned genomes for accelerated evolution of non-coding regions (Pollard et al. 2006; Prabhakar et al.

2006; Haygood et al. 2007; Kim and Pritchard 2007; Bird et al. 2007). Perhaps the two most significant limitations of these previous studies are 1) the lack of a suitable neutral reference with the same mutation rate as each noncoding region of interest (Zhen and Andolfatto 2012); and 2) only regions with the most rapid overall divergence are detected, so that any signature of positive selection can be overpowered by negative selection acting within the same region (Zhen and Andolfatto 2012). Our approach provides a solution to both of these issues. First, by classifying SNVs as neutral only if they have no measurable effect on transcriptional output, we have a reliable neutral reference that is interspersed with non-neutral sites (and thus should be robust to regional variation in mutation rate), in much the same way as synonymous and nonsynonymous sites in protein-coding regions. Second, by distinguishing between up- and down-regulating mutations, we have the ability to detect both positive and negative selection acting simultaneously on different sites within a single enhancer, as we observed for LTV1. Importantly, this positive selection would have been missed by any approach that simply scans for an overall evolutionary rate faster than neutral, because the negative selection on down-regulating mutations (amplified by the fact that there are over twice as many possible down-regulating mutations as up-regulating mutations) actually leads to an overall rate for LTV1 that is *slower* than neutral (The neutral-site divergence of LTV1 across rodents is 115 substitutions / 254 possible neutral SNVs = 45.3%, while the overall divergence of LTV1 across rodents is 352 substitutions / 906 possible SNVs = 38.9%). Only by accounting for the direction of each possible mutation's effect were we able to detect the positive selection that has occurred.

Despite its advantages, there are a number of important limitations of our test. Perhaps most important is the current dearth of saturation mutagenesis data sets that can be used as input for the test (Patwardhan et al. 2009; Melnikov et al. 2012; Patwardhan et al. 2012; Kwasnieski et al. 2012). However these studies are becoming increasingly straightforward to implement (e.g. no longer requiring access to expensive large-scale oligonucleotide synthesis technology (Patwardhan et al. 2012)), making them accessible to any investigators. Moreover the fact that all sites within an enhancer need not be analyzed to detect selection (Supplemental Figure 1) suggests that partial mutagenesis is a viable option. A second caveat is that because *trans*-acting factors can change between species, SNV effects may be species-specific. While this can certainly be an issue at long timescales (e.g. across vertebrates (Ritter et al. 2010; Ariza-Consano et al. 2012)), nearly all human-mouse gene expression divergence has been found to be *cis*-acting (Wilson et al. 2008), making this a minor concern at the even shorter timescales used here. Third, the SNVs are classified according to their transcriptional effects in the livers of mice raised in the laboratory; whether they may have other effects in different tissues/environments is unknown. Additional limitations related to the saturation mutagenesis data include the lack of information on indels, and potential effects of SNVs not captured by the experiment (e.g. mutations that influence enhancer activity in the chromosomal but not the plasmid context, or with effect sizes too small to measure). We expect that most of these limitations will be addressed by more comprehensive saturation mutagenesis studies (perhaps targeted toward the indels or SNV combinations observed in nature) in the near future.

Many extensions to this test are possible. For example, intra-species polymorphism data could be incorporated to allow a McDonald-Kreitman framework to be applied (McDonald and Kreitman 1991), which may allow more sensitive detection of positive selection. In addition, SNV effect sizes could be incorporated to potentially increase the power to detect selection. For example, even when the number of up-regulating SNVs observed in a phylogeny is consistent with the neutral expectation, if they are shifted towards very strongly up-regulating, selection for up-regulation may still be detectable (cf. Moses 2009). Finally, our approach could also be applied to protein-coding regions that have been subjected to saturation mutagenesis (Fowler et al. 2010;

Araya and Fowler 2011; Starita et al. 2013). The framework we have introduced here will likely have many other extensions as well, as our ability to determine the effects of mutations in both coding and noncoding regions continues to evolve.

**Materials and Methods**

*Obtaining present-day enhancer sequences*

      Orthologous mammalian enhancer sequences were identified using the sequences from Patwardhan et al. 2012 (ALDOB (hg19:chr9:104195570-104195828), ECR11 (hg19: chr2:169939082-169939701), and LTV1 (mm9:chr7:29161443- 29161744)) as BLAST query sequences. Using the NCBI Genomes (chromosome) database, we identified the genomic region from each species with the highest sequence identity to the query (with at least 70% identity), and then confirmed its genomic proximity to the putative target gene. If the enhancer sequence was not available in this database, the Whole-Genome Shotgun Contigs (wgs) database was used. In this case, to determine whether the enhancer sequence was located adjacent to the correct gene, several exons of the putative target gene were also input to BLAST to ensure that these sequences mapped to the same contig.

*Ancestral reconstructions*

      Ancestral reconstructions were performed using the Ancestor v1.1 web server, which implements a context-dependent maximum likelihood substitution inference algorithm (Diallo, Makarenkov, and Blanchette 2010). We provided Ancestor v1.1 with alignments of present-day enhancer sequences, including outgroup species not shown in our trees (to improve reconstructions of the most basal nodes; species listed in Supplemental Table 1). Six separate alignments were constructed for each enhancer using different alignment algorithms or combinations of algorithms: 1. PRANK; 2. MUSCLE MSA; 3. tCoffee; 4. ClustalW2; 5. A combination of tcoffee_msa, clustalw_msa, muscle_msa, and clustalw_pair using http://tcoffee.crg.cat/; and 6) a combination of clustalw_pair and lalign_id_pair using http://tcoffee.crg.cat/ (Edgar 2004; Larkin et al. 2007; Löytynoja and Goldman 2010; Di Tommaso

et al. 2011). Default settings were used for all alignments. Alignments were inspected by eye and a small number of poorly aligned regions were manually adjusted (or in one case, the ECR11 alignment using clustalw_pair and lalign_id_pair, excluded). In addition, the Hominoidea (apes and human) contain a LINE element in ECR11, which we removed from our reconstructions (human sequence:

GAAAAATAGATCAATTTGTTCTCACTCATAGGTGGGAATTGAACAATGAGAACACATGGACACAGGAAGGGGAACATCACACATCGGGGCCTGTTGTGGGGTGGGGGGAGGGGGGAGGGATAGCATTAGGAGATATATCTAACGTTAAATGACGTGTTAATGGGAGCAGCACACCAACATGGCACATGTATACATATGTAACAAACTGCATGTTGTGCACATGTACCCTAAAACTTAAAGTATAATAAGAAAAA).

We provided Ancestor v1.1 with an ultrametric phylogeny of all the species included in the alignments (the 'best dates' nexus tree from (Bininda-Emonds et al. 2007)). Branch lengths were rescaled to fall between 0 and 1. For two species missing from this phylogeny we used the most closely related species in the tree as its replacement for our branch length calculations (For Olive Baboon, *Papio anubis*, we substituted *Papio hamadryas*, the Hamadryas Baboon. For the Sumatran Orangutan, *Pongo abelii*, we substituted *Pongo pygmaeus*, the Bornean Orangutan).

For each ancestral node, its six alignments were used to create separate reconstructions, which were then aligned with tCoffee (using default settings). The most frequently observed base in each position was used to generate a consensus, followed by manual curation for ambiguous positions. For a small number of ambiguous positions where there was no consensus, the reconstructions derived from the tCoffee alignments (algorithm #3 above) were given priority. The final consensus reconstructions were used in all subsequent calculations.

*Calculating $K_u/K_n$ and $K_d/K_n$*

$K_u/K_n$ and $K_d/K_n$ were determined by applying expression differences due to single nucleotide variants (SNVs) (Patwardhan et al. 2012) to our ancestral reconstructions to detect evidence of selection (Figure 1). Insertions and deletions (indels) were not included in these calculations, as their impact on transcription was not tested (Patwardhan et al. 2012).

We calculated $K_u$ as the ratio of observed significant ($p < 0.05$, quantifying the probability of the SNV having no effect on transcription (Patwardhan et al. 2012)) up-regulating SNVs divided by the total possible number of significant up-regulating SNVs in each enhancer (Figure 1). An equivalent calculation was performed for $K_d$ and $K_n$ (replacing up-regulating mutations by down-regulating [$p < 0.05$] or neutral [$p > 0.05$]). Varying this p-value threshold (e.g. to 0.01 or 0.1) had little effect on our results (Supplemental Table 3). Patwardhan et al. (2012) tested two independent libraries for LTV1, so we used Fisher's Method to combine the two sets of p-values for use in our calculations.

Fisher's Exact Test was applied to a 2x2 contingency table to determine whether each $K_u/K_n$ or $K_d/K_n$ was significantly different from one (e.g. for $K_u/K_n$, the table columns were up-regulating or neutral, and the rows were observed or not observed in a given branch or phylogeny). For calculations at the level of phylogenies, the counts of each SNV class in each branch were summed.

If our three classes of SNVs (up-regulating, down-regulating, and silent) had different mutation rates, this could affect the estimation of $K_u/K_n$ and $K_d/K_n$. For example, higher mutation rates of neutral SNVs might lead to underestimates of both metrics. To test if this was an issue for the three enhancers studied here, we tabulated the number of transitions and transversions among each class of SNVs within each enhancer (because transitions have a higher mutation rate than transversions). We found no significant difference for any of them ($p = 0.20, 0.47,$ and $0.88$ for ALDOB, LTV1, and ECR11 respectively). For enhancers where there is a difference in mutation rates between classes, this could be easily incorporated by adjusting $K_u$, $K_d$, and $K_n$ appropriately.

**Acknowledgements**

We would like to thank the members of the Fraser Lab for helpful discussions. This work was supported by the National Institute of General Medical Sciences (grant number 1R01GM097171-01A1). JDS is supported by a Stanford Graduate Fellowship and a Genentech Graduate Fellowship. HBF is an Alfred P. Sloan Fellow and a Pew Scholar in the Biomedical Sciences.

**Figure Legends**

**Figure 1. Outline of our approach.** Using expression maps of enhancers and the effect of every single possible single nucleotide variant (SNV) combined with present-day sequences and ancestral reconstructions, our method estimates $K_u/K_n$ and $K_d/K_n$ to detect evidence of selection on cis-regulatory elements. $K_u/K_n$ or $K_d/K_n$ values significantly greater than one signify positive selection, while values significantly less than one indicate negative selection.

**Figure 2. Selection on LTV1. A.** Phylogenetic tree of the rodent species for which LTV1 enhancer sequences were analyzed. Ancestral nodes are labeled L1-L7, and branches are labeled A-M. **B.** $K_u/K_n$ and **C.** $K_d/K_n$ values are plotted for each branch of the phylogenetic tree. Below each bar is the number of up- or down-regulating (top row) and neutral (bottom row) substitutions inferred for that branch. Asterisks mark branch-specific $K_u/K_n$ or $K_d/K_n$ values that differed significantly from neutral (Fisher's Exact $p < 0.05$).

**Figure 3. Selection on ALDOB. A.** Phylogenetic tree of the primate species for which ALDOB enhancer sequences were analyzed. Note that A8 appears twice as the reconstruction did not differ between these two nodes. Ancestral nodes are labeled A1-A10, and branches are labeled A-T. **B.** $K_u/K_n$ and **C.** $K_d/K_n$, as in Figure 2.

**Figure 4. Selection on ECR11.** A) Phylogenetic tree of the primate species for which ECR11 enhancer sequences were analyzed. Note that E3 appears twice as the reconstruction did not differ between these two nodes. Ancestral nodes are labeled E1-E11, and branches are labeled A-V. **B.** $K_u/K_n$ and **C.** $K_d/K_n$, as in Figure 2.

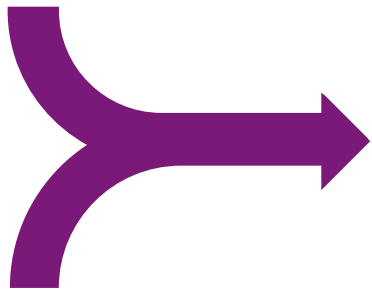

# A LTV1

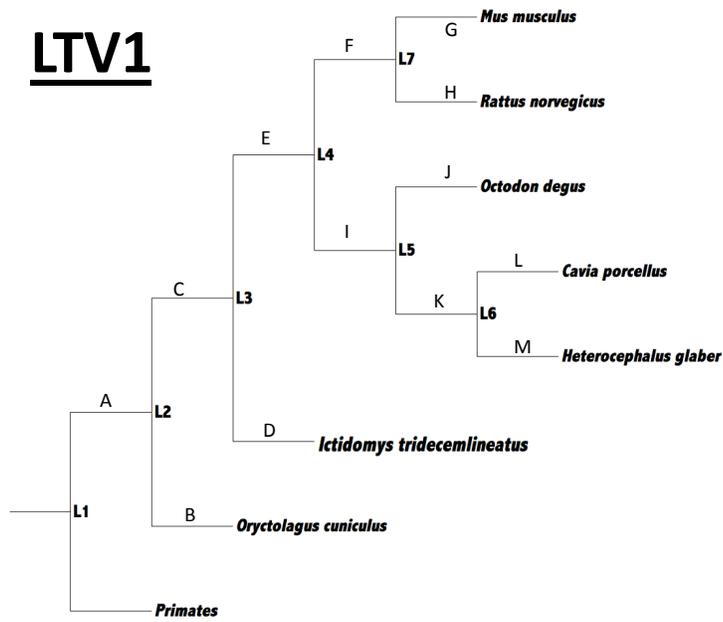

# B

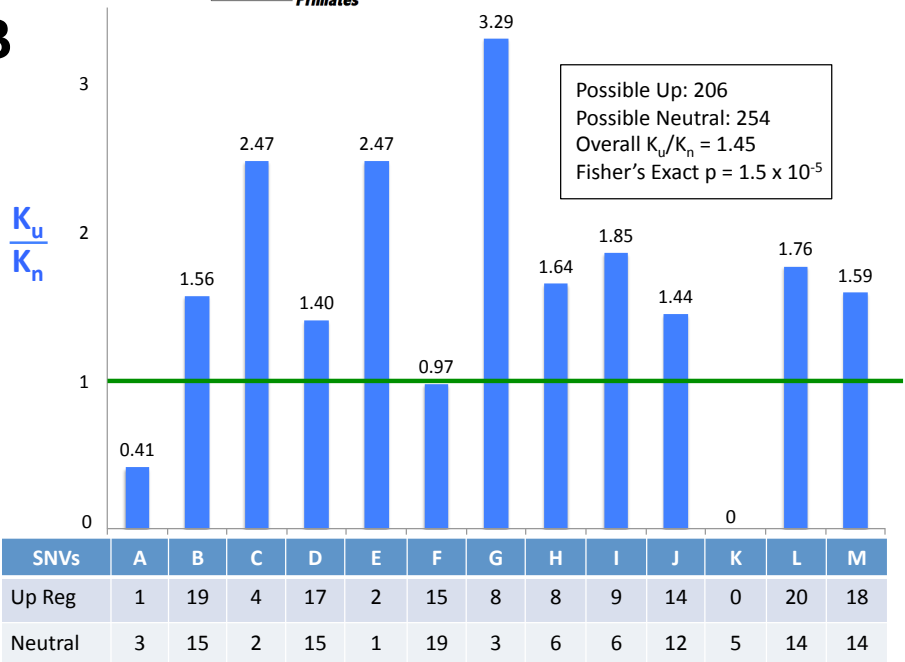

Possible Up: 206
Possible Neutral: 254
Overall $K_u/K_n$ = 1.45
Fisher's Exact p = 1.5 x $10^{-5}$

| SNVs | A | B | C | D | E | F | G | H | I | J | K | L | M |
|---|---|---|---|---|---|---|---|---|---|---|---|---|---|
| Up Reg | 1 | 19 | 4 | 17 | 2 | 15 | 8 | 8 | 9 | 14 | 0 | 20 | 18 |
| Neutral | 3 | 15 | 2 | 15 | 1 | 19 | 3 | 6 | 6 | 12 | 5 | 14 | 14 |

# C

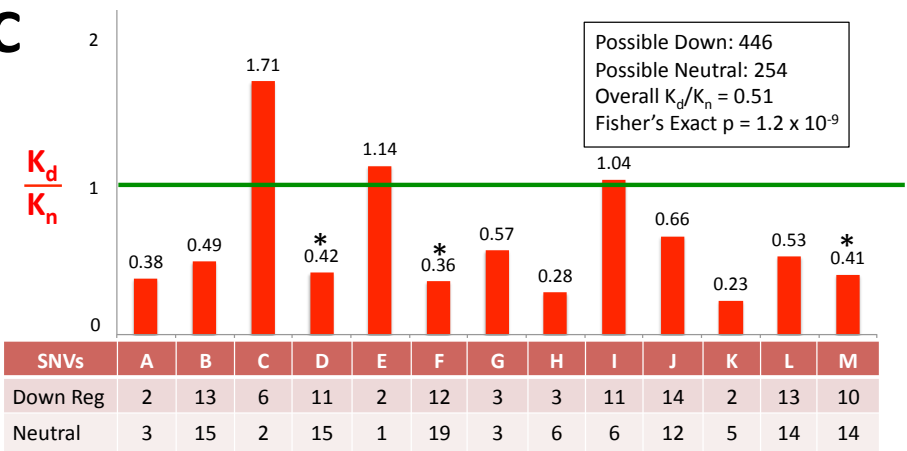

Possible Down: 446
Possible Neutral: 254
Overall $K_d/K_n$ = 0.51
Fisher's Exact p = 1.2 x $10^{-9}$

| SNVs | A | B | C | D | E | F | G | H | I | J | K | L | M |
|---|---|---|---|---|---|---|---|---|---|---|---|---|---|
| Down Reg | 2 | 13 | 6 | 11 | 2 | 12 | 3 | 3 | 11 | 14 | 2 | 13 | 10 |
| Neutral | 3 | 15 | 2 | 15 | 1 | 19 | 3 | 6 | 6 | 12 | 5 | 14 | 14 |

# A ALDOB

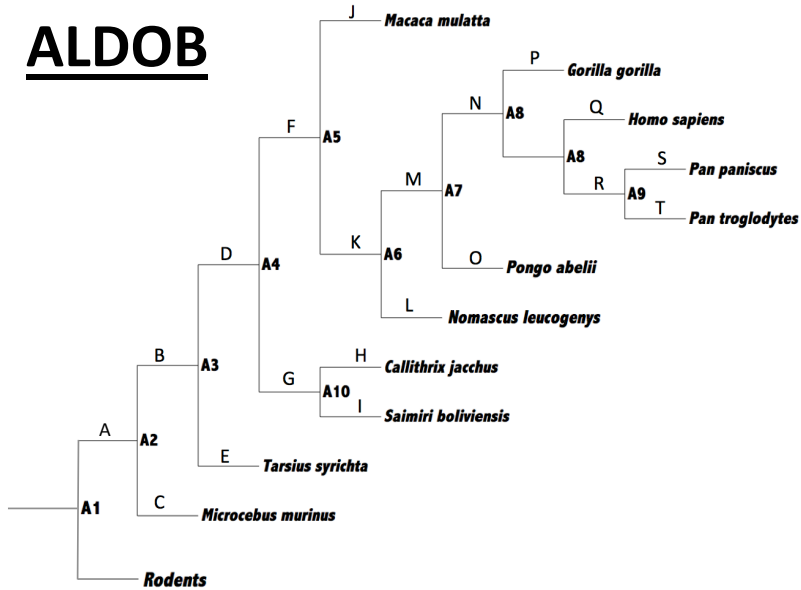

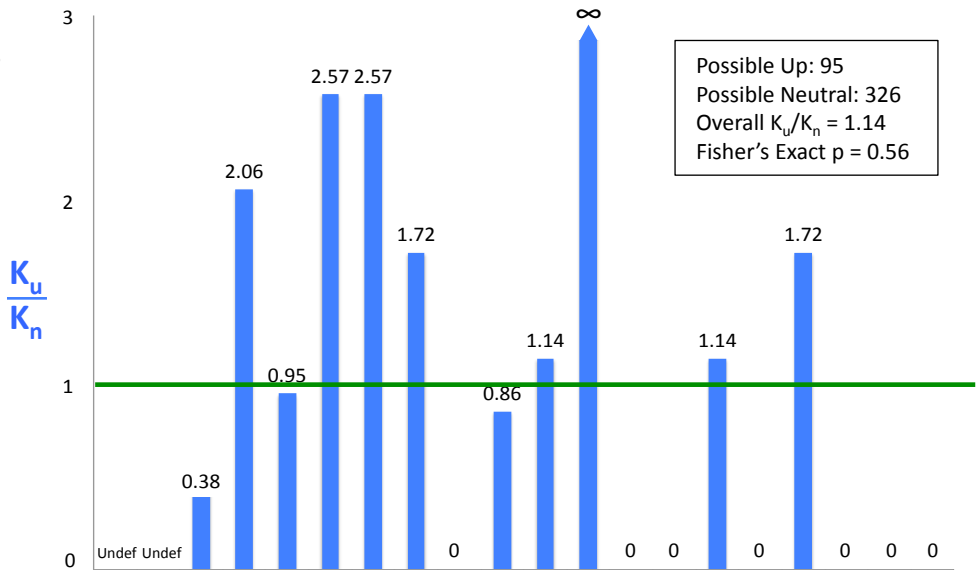

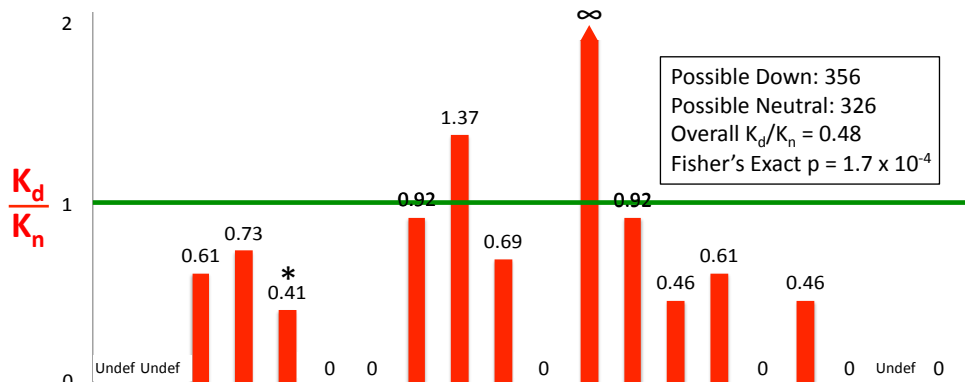

# A ECR11

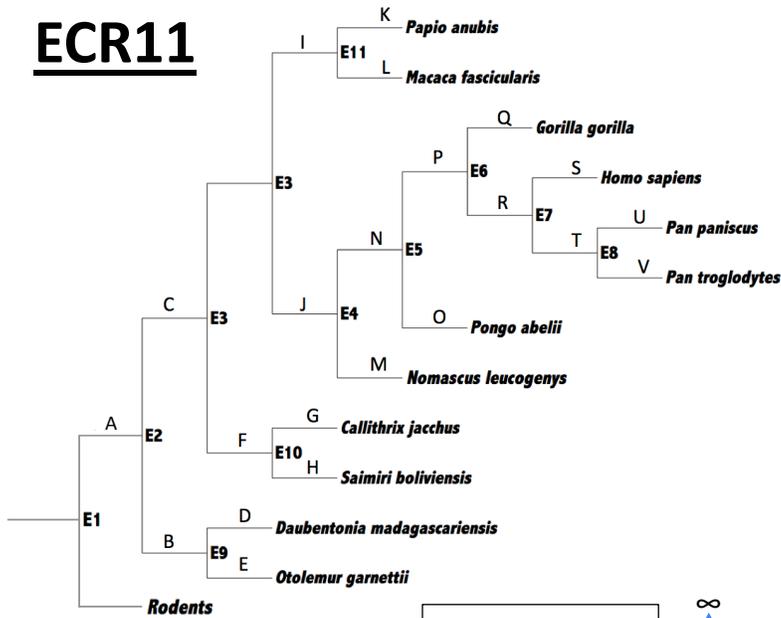

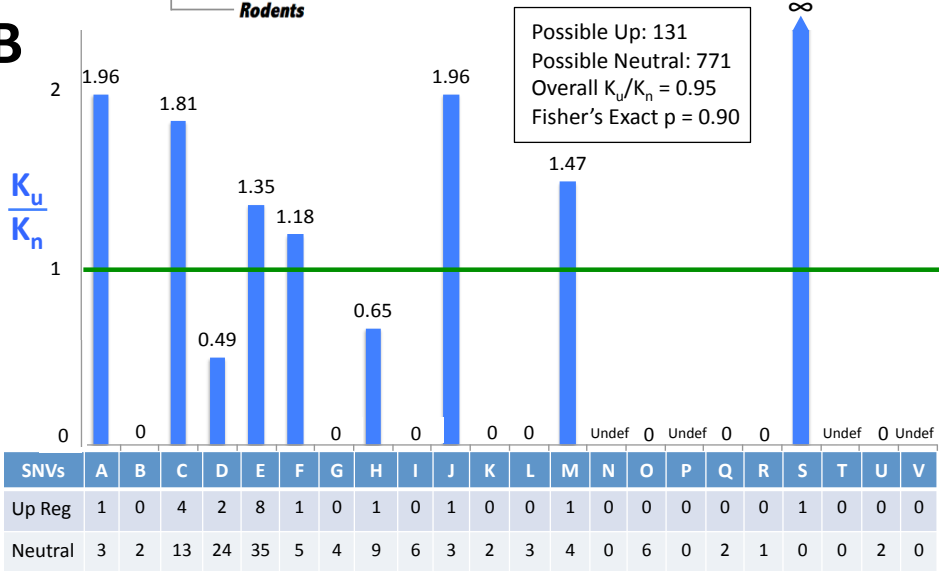

Possible Up: 131
Possible Neutral: 771
Overall $K_u/K_n$ = 0.95
Fisher's Exact p = 0.90

| SNVs | A | B | C | D | E | F | G | H | I | J | K | L | M | N | O | P | Q | R | S | T | U | V |
|---|---|---|---|---|---|---|---|---|---|---|---|---|---|---|---|---|---|---|---|---|---|---|
| Up Reg | 1 | 0 | 4 | 2 | 3 | 1 | 0 | 1 | 0 | 1 | 0 | 1 | 0 | 0 | 0 | 0 | 0 | 0 | 1 | 0 | 0 | 0 |
| Neutral | 3 | 2 | 13 | 24 | 35 | 5 | 4 | 9 | 6 | 3 | 2 | 3 | 4 | 0 | 6 | 0 | 2 | 1 | 0 | 0 | 2 | 0 |

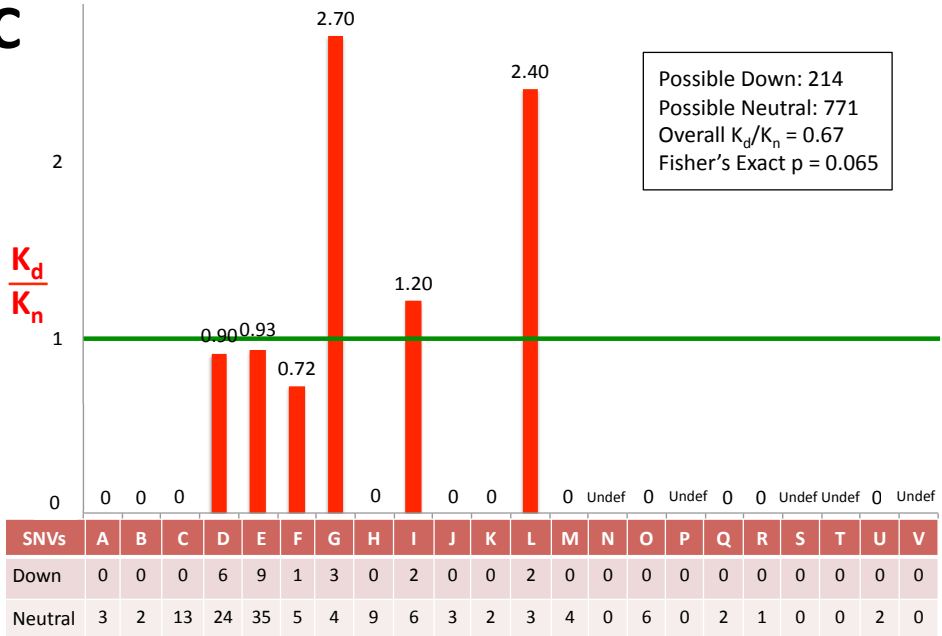

Possible Down: 214
Possible Neutral: 771
Overall $K_d/K_n$ = 0.67
Fisher's Exact p = 0.065

| SNVs | A | B | C | D | E | F | G | H | I | J | K | L | M | N | O | P | Q | R | S | T | U | V |
|---|---|---|---|---|---|---|---|---|---|---|---|---|---|---|---|---|---|---|---|---|---|---|
| Down | 0 | 0 | 0 | 5 | 9 | 1 | 3 | 0 | 2 | 0 | 0 | 2 | 0 | 0 | 0 | 0 | 0 | 0 | 0 | 0 | 0 | 0 |
| Neutral | 3 | 2 | 13 | 24 | 35 | 5 | 4 | 9 | 6 | 3 | 2 | 3 | 4 | 0 | 6 | 0 | 2 | 1 | 0 | 0 | 2 | 0 |